\documentclass[12pt]{article}

\input{epsf}
\usepackage{amsfonts}
\usepackage{amsmath}
\usepackage{amssymb}
\usepackage{latexsym}
\usepackage[dvips]{graphicx}
\usepackage{color}

\newcommand{\be}{\begin{equation}}
\newcommand{\ee}{\end{equation}}
\newcommand{\bea}{\begin{eqnarray}}
\newcommand{\eea}{\end{eqnarray}}

\newcommand{\hphi}{\hat{\phi}}
\newcommand{\hpi}{\hat{\Pi}}

\numberwithin{equation}{section}

\begin{document}
\baselineskip 20pt
\begin{titlepage}
\begin{center}
{\Large {\bf Quantum dynamics of the\\[1ex]
Einstein-Rosen wormhole throat}}
\vspace{1cm} 


\renewcommand{\baselinestretch}{1}
{\bf
Gabor Kunstatter${}^\dagger$, 
Jorma Louko${}^\sharp$
and
Ari Peltola${}^\dagger$
\\}
\vspace*{0.7cm}
{\sl
${}^\dagger$ 
Department of Physics and\\
Winnipeg Institute for Theoretical Physics 
The University of Winnipeg\\
515 Portage Avenune, 
Winnipeg, Manitoba, Canada R3B 2E9\\
{[e-mail: g.kunstatter@uwinnipeg.ca, peltolanari@gmail.com]}\\[5pt]
}
{\sl
${}^\sharp$ 
School of Mathematical Sciences,
University of Nottingham\\
Nottingham NG7 2RD, United Kingdom\\
{[e-mail: jorma.louko@nottingham.ac.uk]}\\ [5pt]
}
\vspace{2ex}
{\bf Abstract}
\end{center}
We consider the polymer quantization of the Einstein wormhole throat theory for an eternal Schwarzschild black hole. We  numerically solve the difference equation describing the quantum evolution of an initially Gaussian, semi-classical wave packet. As expected from previous work on loop quantum cosmology, the wave packet remains semi-classical until it nears the classical singularity at which point it enters a quantum regime in which the fluctuations become large. The expectation value of the radius reaches a minimum as the wave packet is reflected from the origin and emerges to form a near Gaussian but asymmetrical semi-classical state at late times. The value of the minimum depends in a non-trivial way on the initial mass/energy of the pulse, its width and the polymerization scale. For wave packets that are sufficiently narrow near the bounce, the semi-classical bounce radius is obtained. Although the numerics become 
difficult to control 
in this limit, we argue that for pulses of finite width the bounce persists as the polymerization scale goes to zero, suggesting that in this model the loop quantum gravity effects mimicked by polymer quantization do not play a crucial role in the quantum bounce.
\vfill \hfill 
Revised December 2010
\vfill \hfill 
Published in Phys.\ Rev.\  D {\bf 83}, 044022 (2011) 

\end{titlepage}

\section{Introduction}

There exists a large body of evidence supporting the notion that loop quantum gravity has the potential to resolve the singularities that plague classical cosmology \cite{lqc1,lqc2,lqc3,Ashtekar:2007em,corichi08} and black hole physics \cite{ashtekar05,modesto06,boehmer07,pullin08,nelson08,pk09}. Most of this work uses 
the polymer representation of quantum mechanics \cite{afw,halvorson} as a way of mimicking the features of a more complete theory of loop quantum gravity. It is important to note that polymer quantization is a viable quantization scheme in its own right, one that is unitarily inequivalent to the usual Schr\"odinger quantization. Moreover, one of its defining features is a fundamental discreteness at some intrinsic microscopic length scale. It is plausible that an underlying discreteness at short distance scales is a general feature of any theory of quantum gravity, including loop quantum gravity, so that the qualitative features that emerge from polymer quantization may give clues regarding the short distance effects of quantum gravity, irrespective of its detailed microscopic structure.
In this spirit, polymer quantization has been applied
to a variety of different quantum mechanical systems including the harmonic
oscillator~\cite{afw}, Coulomb potential \cite{hlw} and $1/r^2$
potential~\cite{klz}. An important question that arises in the context of singular potentials concerns the extent to which polymer quantization provides a mechanism for singularity resolution that is fundamentally 
distinct from Schr\"odinger quantization. 

We focus on a model for the interior of an eternal Schwarzschild black hole that was first developed in the quantum setting by Louko and M\"akel\"a \cite{Louko:1996md}.  The dynamics is that of the minimum radius of the throat of the Einstein-Rosen bridge as a function of the proper time of a comoving observer at the throat. This model has two  advantages over other models of black hole interiors that have appeared in the recent literature. First, it is fully reduced and describes the evolution of an eternal Schwarzschild black hole in terms of geometrical invariants. Second, it produces relatively simple equations at the classical and quantum level that are amenable to both analytic and numerical analysis. One disadvantage of the model is that it is difficult to reconstruct the full black hole quantum corrected spacetime from the fully reduced model, as was done in \cite{pullin08,pk09}. 

The current work is a continuation of \cite{kpl} in which the full polymer quantized throat theory was constructed and the energy spectrum was obtained using a numerical 
shooting method. 
In the present paper we numerically evolve an initially Gaussian wave packet 
in the polymer quantized throat theory. 
The initial packet is chosen to be semi-classical in the sense that its width is large compared to the polymer scale but small compared to the only other scale in the problem, namely the ADM mass. To the best of our knowledge, the analogous calculation for this model using Schr\"odinger quantization has not been done but our results can be compared to the semi-classical limit of the Schr\"odinger throat theory obtained in Ref.~\cite{Louko:1996md}. 

The paper is organized as follows: In Section 2 we review the throat theory that we use to describe the classical dynamics of the black hole interior. Section 3 summarizes previous results on the effective polymer dynamics, which yield the semi-classical spectrum and time evolution of the throat. Section 4 describes the full polymer theory as well as our numerical methods and results. Section 5 closes with a summary, conclusions and prospects for future work.


\section{Classical theory}
\label{sec:class}

We consider a Hamiltonian system with a 
two-dimensional phase space and the Hamiltonian 
\be 
\label{eq:classH1} 
H = \frac{1}{2}\left(\frac{p^2}{r}+r\right),
\ee
where the configuration variable $r$ takes positive values and $p$ is the conjugate momentum. The equations of motion reduce to 
\be 
\label{eq:geodesic} 
\dot{r}^2=\frac{2M}{r}-1,
\ee
where $M$ is the conserved value of $H$ and the overdot denotes 
derivative with respect to the time~$t$. 
Note that $M$ is necessarily positive. 

\begin{figure}[htbp!]
\begin{center}
\includegraphics[scale=0.5]{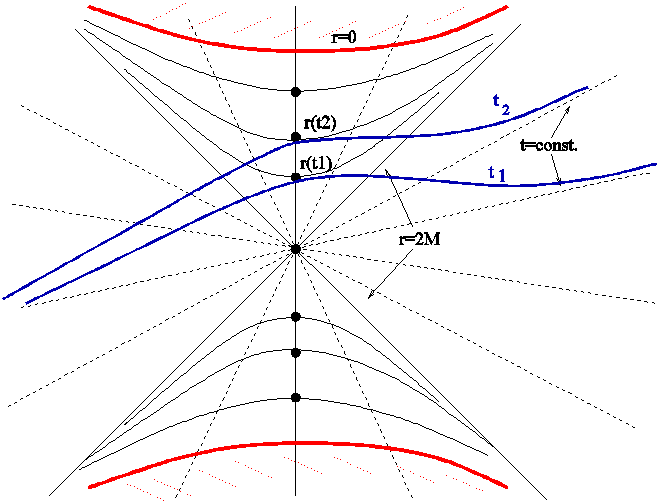}
\caption{Schematic of the geometrodynamics of the time evolution of the throat of the Einstein-Rosen bridge as a function of proper time of a comoving observer}
\label{fig:kruskal}
\hfill

\hfill


\includegraphics[scale=0.6]{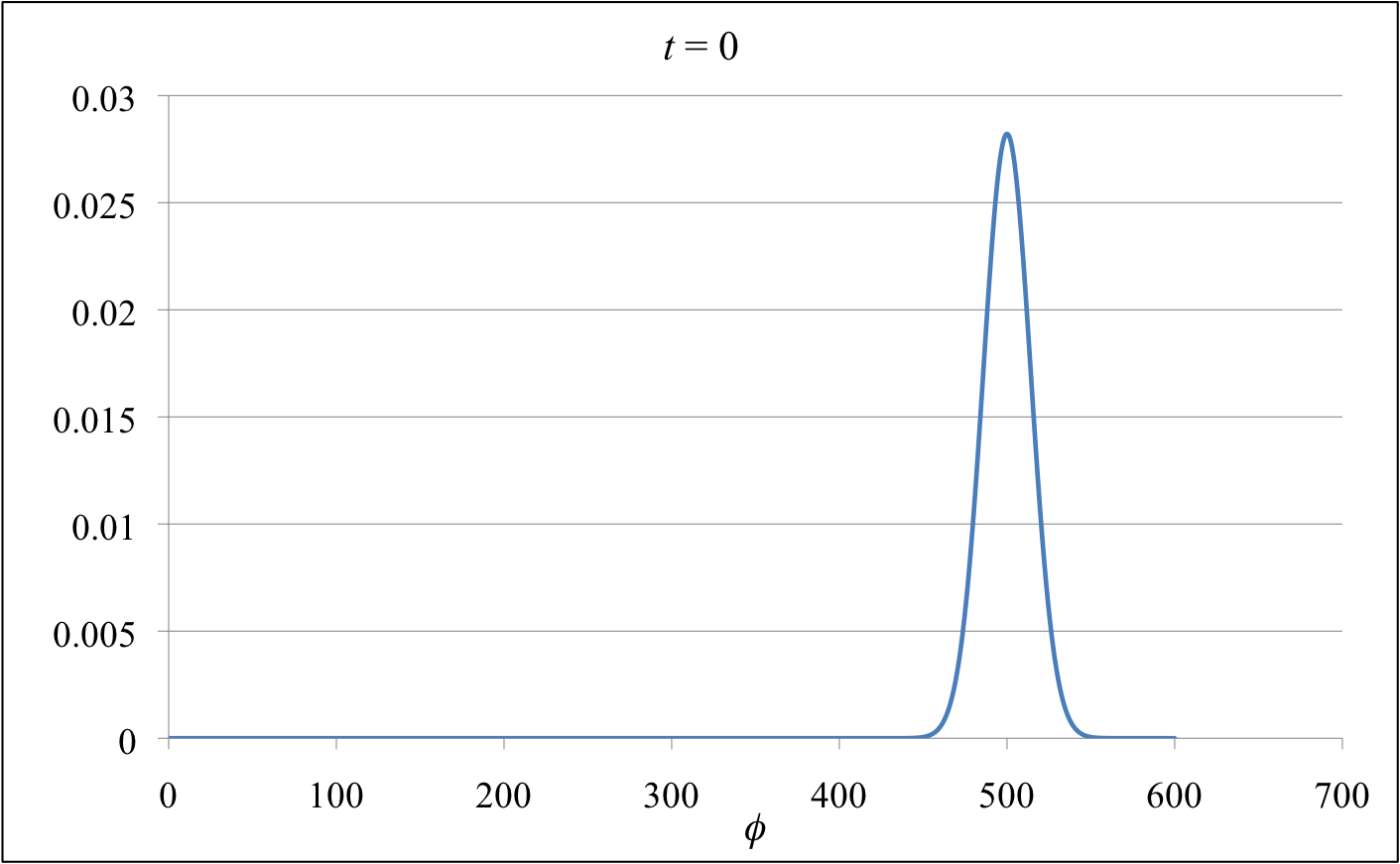}
\caption{Time evolution of wave packet, $t=0$. In this figure $\mu=1$, $\delta = 0.05$, $\phi_0=500$ and $\pi_0=-1\times 10^{-7}\approx 0$.}
\label{fig:phi500_t0}
\end{center}
\end{figure}

The equation of motion \eqref{eq:geodesic} is identical to the equation of a radial timelike geodesic that passes through the bifurcation two-sphere on a Kruskal manifold of mass~$M$, provided $r$ is identified with the area-radius of the two-sphere and $t$ is identified with the proper time on the geodesic~\cite{friedman,redmount}. This suggests that the Hamiltonian \eqref{eq:classH1} could be interpreted as the Hamiltonian of the Einstein-Rosen wormhole throat. It was shown in \cite{Louko:1996md} that this interpretation follows from a direct derivation of the Hamiltonian \eqref{eq:classH1} from the geometrodynamical action of general relativity, after reducing the theory to spherical symmetry and anchoring the evolution of the spacelike hypersurfaces into the two asymptotically flat spacelike infinities of the Kruskal manifold in a particular fashion~\cite{Kuchar:1994zk}. To ensure that the value of the Hamiltonian equals the Schwarzschild mass, the evolution at one infinity is taken to proceed at unit rate in the asymptotic Minkowski time~$t$, while the evolution at the other end is frozen. To express the Hamiltonian in terms of the wormhole throat radius, the hypersurfaces are chosen to coincide with those of the Einstein-Rosen wormhole near the wormhole throat, and the time evolution is chosen so that 
$t$ coincides with the proper time at the wormhole throat. Under these conditions the Hamiltonian duly takes the form~\eqref{eq:classH1}, with the variable $r$ being the area-radius of the wormhole throat~\cite{Louko:1996md}. A~sketch of the foliation is shown in Figure~\ref{fig:kruskal}. 

To polymer quantize the theory, a choice must be made for the variable to be polymerized, and this choice can affect the qualitative properties of the quantum theory~\cite{lqc2,lqc3}. 
For reasons described in \cite{kpl} we work with the 
canonical chart $(\phi,\Pi)$, where
{\setlength\arraycolsep{2pt}
\begin{subequations} 
\label{eq:polv}
\bea \phi&=&r^{2}, 
\\ 
\Pi &=& \frac12 \frac{p}{ r}  \ , 
\eea
\end{subequations}}
with $0 < \phi < \infty$ and $-\infty < \Pi < \infty$. 
The Hamiltonian takes the form 
\be 
\label{eq:classH2}
H = \frac{\phi^{1/2}}{2} 
\left( 4\Pi^2 + 1 \right) \,  . 
\ee
We work in Planck units, $G=c=\hbar=1$. 
The conversion to geometric units, in which $G=c=1$ but $[\hbar] =
(\text{length})^2$, is discussed in~\cite{kpl}.


\section{Effective polymer theory}
\label{sec:effpolytheory}

As discussed in more detail 
in \cite{pullin08,husain06}, 
a semiclassical, or effective, polymer approximation is obtained by replacing $\Pi\mapsto\sin (\mu\Pi)/\mu$ in the classical 
Hamiltonian, with the result 
\be H_\text{eff} = \frac{\phi^{1/2}}{2} \bigg(\frac{4\sin^2(\mu\Pi)}{\mu^2}+1\bigg),
\ee
where $\mu$ is the polymerization scale. The resulting equations of motion are thought to approximate the time evolution of the expectation value of $x$ and $p$ for semi-classical states. 
These equations read
\begin{align} \dot{\phi} &= \{\phi,H_\text{eff}\} = \frac{4\phi^{1/2}\sin (\mu\Pi)\cos(\mu\Pi)}{\mu},\\[4pt]
\dot{\Pi} &= \{\Pi,H_\text{eff}\} = - \frac{M}{2\phi} , 
\end{align}
where $M$ denotes the constant value of $H_\text{eff}$ on a solution. 

It 
follows that $\phi$ has two turning points where $\dot{\phi}=0$:
\begin{align} \sqrt{\phi_+} &= 2M,
\label{eq:phi_+} \\ 
\sqrt{\phi_-} &= 2M\bigg(\frac{\mu^2}{4+\mu^2}\bigg).
\label{eq:phi_-}
\end{align}
Solving for $\dot{\phi}$ in terms of $\phi$, $\mu$ and $M$ gives
\begin{align} \dot{\phi}^2 &= 4\phi\bigg[\frac{2M}{\phi^{1/2}}-1\bigg]\bigg[1-\frac{\mu^2}{4}\bigg(\frac{2M}{\phi^{1/2}}-1\bigg)\bigg] \nonumber\\[4pt]
&= -\big(4+\mu^2 \big)\Big[ \phi - (\phi_+^{1/2}+\phi_-^{1/2})\phi^{1/2}+\phi_+^{1/2}\phi_-^{1/2}\Big].
\end{align}
By integration we obtain
\begin{align} \label{eq:t_phi} t=\frac{2}{\sqrt{4+\mu^2}}\bigg[&\sqrt{(2M-\phi^{1/2})(\phi^{1/2}-2M\kappa)}\nonumber \\ &+M(1+\kappa)\arcsin \bigg(\frac{2M(1+\kappa)-2\phi^{1/2}}{2M(1-\kappa)}\bigg)\bigg] +C,
\end{align}
where $\kappa = \mu^2/(4+\mu^2)$. This equation gives implicitly the trajectory $\phi(t)$. 
Equations (\ref{eq:phi_+}), (\ref{eq:phi_-}) and (\ref{eq:t_phi}) are the key ingredients from the effective theory that we will need for comparison with the time evolution in the full polymer theory.

\section{Full polymer theory}

Here we summarize only the features of the quantum theory that are needed for the subsequent calculation. We refer the reader to \cite{afw, halvorson, kpl} and references therein for further details. 

\subsection{Operators}

We take the polymer Hilbert space $\mathcal{H}$ to have the 
orthonormal basis 
$\left\{|m\mu\rangle \mid m \in \mathbb{Z} \right\}$
of eigenstates of the operator~$\hphi$, 
\be 
\hphi |m\mu\rangle = m\mu\, |m\mu\rangle , 
\ee
where $\mu$ is the polymerization scale 
as in Section~\ref{sec:effpolytheory}. 
The orthonormality relation reads 
\be 
\langle m\mu | m'\mu \rangle = \delta_{m,m'}.
\ee 
The momentum operator $\hat{\Pi}$
is defined by 
\be 
\label{eq:momentum}
\hat{\Pi}=\frac{1}{2i\mu}\big( \hat{U}_{\mu}^\dagger-\hat{U}_{\mu}\big),
\ee
where $\hat{U}_{\mu}$ is the finite translation operator, 
\be 
\hat{U}_{\mu} | m \mu \rangle =|(m+1)\mu\rangle . 
\label{eq:U_hat}
\ee
We note in passing that equations 
(\ref{eq:momentum}) and (\ref{eq:U_hat}) correct a sign error 
in the definition of $\hat{\Pi}$ in~\cite{kpl} (see~\cite{afw}); 
this error does however not affect the conclusions in \cite{kpl} as 
the polymer Hamiltonian (equation (\ref{eq:H_phi}) below) is quadratic 
in~$\hat{\Pi}$. 

We order the polymer Hamiltonian operator as \cite{kpl}
\be 
\hat{H}_\text{pol} 
= \frac{1}{2}\Big(4\hpi\hphi^{1/2}\hpi+\hphi^{1/2}\Big) .
\label{eq:H_phi}
\ee
As $\hphi$ is not positive definite, 
the square roots in \eqref{eq:H_phi} 
are not a priori defined. 
Following~\cite{kpl}, we define these square roots as 
\be 
\label{eq:hphi} 
\hphi^{1/2} |m\mu\rangle := |m\mu|^{1/2}|m\mu\rangle . 
\ee
The action of $\hat{H}_\text{pol}$ on the basis states then reads 
\begin{align} 
\label{eq:Hpol} 
\hat{H}_\text{pol} |m\mu\rangle = \frac{1}{2\mu^{3/2}} 
\bigg[ &\,\Big( |m+1|^{1/2} + |m-1|^{1/2} + \mu^2 |m|^{1/2} \Big) |m\mu\rangle \nonumber\\
& - |m+1|^{1/2} |(m+2)\mu\rangle - |m-1|^{1/2}  |(m-2)\mu\rangle \bigg] . 
\end{align}

To solve Schr\"odinger's equation, 
\be 
\label{eq:Schrodinger} 
i\frac{\partial}{\partial t} \Psi (t) = \hat{H}_\text{pol} \Psi (t) , 
\ee
we write 
\be \label{eq:gen_state} 
\Psi (t) = \frac{1}{C} \sum_{m=-\infty}^{\infty} \big[f_m(t)+ig_m(t)\big] |m\mu\rangle ,
\ee
where $f_m$ and $g_m$ are real-valued functions of $t$ and 
$C$ is a 
real-valued 
normalization constant. 
Eq.~(\ref{eq:Schrodinger}) then reduces into 
a system of coupled first-order 
differential equations for the coefficient functions $\left\{f_m\right\}$ 
and~$\left\{g_m\right\}$.

\subsection{Initial state}
We take the initial state to be 
\be \label{eq:initial}\Psi (0) = |\phi_0,\pi_0\rangle = \frac{1}{C} \sum_{m=-\infty}^{\infty} e^{-\delta^2(m\mu-\phi_0)^2/(2\mu^2)}\,\, e^{i\pi_0 m\mu} |m\mu\rangle ,
\ee
where $C$ is the normalization constant and the parameter $\delta$ controls the width of the state. Using the notation above,
\be \Psi (0) = \frac{1}{C} \sum_{m=-\infty}^{\infty} \big[f_m(0)+ig_m(0)\big] |m\mu\rangle ,
\ee
with the identifications 
\begin{align} f_m(0) &= e^{-\delta^2(m\mu-\phi_0)^2/(2\mu^2)} \sin (\pi_0m\mu), \\  g_m(0) &= e^{-\delta^2(m\mu-\phi_0)^2/(2\mu^2)} \cos (\pi_0m\mu),
\end{align}
This state is a generalization of a coherent state of quantized harmonic oscillator: 
it 
is peaked at the classical phase space point $(\phi_0,\pi_0)$ \cite{afw,husain06}. 
Note that in the present context 
the state 
can only be considered semi-classical when 
its width 
is large compared to the lattice spacing~$\mu$. To leading order, 
\be
(\Delta \phi) = \frac{\mu}{\sqrt{2}\delta} , 
\ee
where $\Delta \phi$ is the standard deviation of~$\phi$. 
Hence the condition for the state being semiclassical is that $\sqrt{2}\delta \ll1$.

We 
start the time evolution when $\langle\phi\rangle$ is near the outer turning point, i.e. the horizon, which requires that $\pi_0\sim 0$. Numerically this means making $\pi_0$ so small that its further decrease does not change the results up to a desired accuracy. The relevant initial parameters are therefore $\phi_0$ and $\delta$.

\subsection{Expectation values}

We have set up the polymer quantum theory so 
that the spectrum of $\hphi$ consists of all 
integer multiples of the polymer scale~$\mu$, 
including the negative integer multiples.  
This enabled us to define the momentum operator 
$\hat{\Pi}$ \eqref{eq:momentum}
and the Hamiltonian operator 
$\hat{H}_\text{pol}$ \eqref{eq:H_phi} 
in terms of the translation operator 
$\hat{U}_\mu$~\eqref{eq:U_hat}. 
As $\hat{H}_\text{pol}$ commutes with the parity operator, 
$|m\mu\rangle \mapsto |(-m)\mu\rangle$, the quantum theory  
decomposes into superselection sectors that consist respectively 
of symmetric and antisymmetric states under parity. 
As the classical theory satisfies 
$0 < \phi < \infty$, each of the two 
superselection sectors can be regarded as a distinct quantization of the classical theory: 
in the limit of small~$\mu$, numerical evidence indeed indicates that the spectra in the two sectors reduce respectively to those of the Schr\"odinger quantized theory with a corresponding boundary condition~\cite{kpl}. 
The emergence of the symmetric and antisymmetric superselection sectors is qualitatively similar to what was found the Coulomb potential in \cite{hlw} and for the inverse square potential in~\cite{klz}.\footnote{Construction  of polymer equivalents of more general Schr\"odinger boundary conditions is currently under investigation~\cite{kl10}.} 

For computations in each of the superselection 
sectors, we may define the `physical' states $\psi (t)$ 
by excluding the negative values of~$m$, 
\be 
\psi (t) := \Psi (t) 
- \frac{1}{C} 
\sum_{m=-\infty}^{-1} \big[f_m(t)+ig_m(t)\big] |m\mu\rangle , 
\ee
and we may normalize the 
inner product of the states $\psi^{(1)}$ and $\psi^{(2)}$ 
to take the form 
\begin{align}
\bigl\langle \psi^{(1)} \big| \psi^{(2)}\bigr\rangle 
&= 
\frac{1}{C^{(1)} \, C^{(2)}}
\Biggl[ 
\frac{1}{2} 
\big(f_0^{(1)} - i g_0^{(1)}\big)
\big(f_0^{(2)} + i g_0^{(2)}\big)
\notag
\\[1ex]
&
\hspace{15ex}
+ \sum_{m=1}^\infty 
\big(f_m^{(1)} - i g_m^{(1)}\big)
\big(f_m^{(2)} + i g_m^{(2)}\big)
\Biggr] . 
\end{align}
A normalized state then satisfies 
\be 
C^2 = \frac{1}{2} \big(f_0^2+g_0^2\big) 
+ \sum_{m=1}^\infty \big(f_m^2+g_m^2\big).
\ee
The expectation value $\langle\phi\rangle$
and the variance $\Delta\phi$ take the form 
\begin{align}
\langle\phi\rangle 
&=
\langle\psi | \hphi |\psi\rangle =\frac{\mu}{C^2} \sum_{m=1}^\infty m\big(f_m^2+g_m^2\big),
\\
(\Delta\phi)^2 
&= 
\frac{1}{C^2}\bigg[\frac{1}{2}\langle\phi\rangle^2\big(f_0^2+g_0^2\big) + \sum_{m=1}^\infty \big(m\mu-\langle\phi\rangle\big)^2 \big(f_m^2+g_m^2\big) \bigg].
\end{align}

\subsection{Numerical methods}
To obtain the coefficients $f_m$ and $g_m$ at a given time instant $t$, we solve the Schr\"odinger equation (\ref{eq:Schrodinger}) using the fourth order Runge-Kutta method. Using Eq.\ (\ref{eq:Hpol}) we obtain
\begin{align} \frac{\partial f_m}{\partial t} +i\frac{\partial g_m}{\partial t} = F_m \big[ \vec g \big] - i F_m  \big[ \vec f \big],
\end{align}
where we have arranged the coefficients $f_m$ and $g_m$ into the vectors $\vec f$ and $\vec g$ and we have defined the functional $F_m \big[ \vec f \big]$ as
\begin{align}  F_m \big[\vec f \big] = \frac{1}{2\mu^{3/2}} \bigg[ &\,\Big( |m+1|^{1/2} + |m-1|^{1/2} + \mu^2 |m|^{1/2} \Big)  f_m \nonumber\\
& - |m+1|^{1/2} f_{m+2} - |m-1|^{1/2}  f_{m-2} \bigg] . 
\end{align}

In the fourth order Runge-Kutta method, the time evolution of the coefficients $f_m$ and $g_m$ is given by the equations
\begin{align} f_m (t+\Delta t) &= f_m (t) + \frac{\Delta t}{6} \Big[k_1^{(m)}  +2k_2^{(m)} +2 k_3^{(m)} +k_4^{(m)} \Big], \\ g_m (t+\Delta t) &= g_m (t) - \frac{\Delta t}{6} \Big[h_1^{(m)}  +2h_2^{(m)} +2 h_3^{(m)} +h_4^{(m)} \Big] ,
\end{align}
where 
\begin{subequations}
\begin{align} k_1^{(m)} &= F_m \big[\vec g\big], \\ k_2^{(m)} &= F_m \big[\vec g +\frac{\Delta t}{2} \vec k_1 \big], \\ k_3^{(m)} &= F_m \big[\vec g +\frac{\Delta t}{2} \vec k_2 \big], \\ k_4^{(m)} &= F_m \big[\vec g +\Delta t\, \vec k_3 \big],
\end{align}
\end{subequations}
and 
\begin{subequations}
\begin{align} h_1^{(m)} &= F_m \big[\vec f\big], \\ h_2^{(m)} &= F_m \big[\vec f +\frac{\Delta t}{2} \vec h_1 \big], \\ h_3^{(m)} &= F_m \big[\vec f +\frac{\Delta t}{2} \vec h_2 \big], \\ h_4^{(m)} &= F_m \big[\vec f +\Delta t\, \vec h_3 \big].
\end{align}
\end{subequations}
To calculate the coefficients $f_m$ and $g_m$ at the next time step $t+\Delta t$, one therefore needs information from the neighboring lattice points at the time instant $t$. For instance, to obtain $f_m(t+\Delta t)$ one needs to know the elements of $\vec g(t)$ up to the lattice points $m\pm 8$. This creates a boundary problem which is dealt with as follows.

The boundary conditions at the origin pose no problem, as explained in the previous section. We can impose either 
symmetric or anti-symmetric boundary conditions
and evaluate the relevant terms in the Runge-Kutta method for both positive and negative~$m$. The results exhibited below used symmetric boundary conditions, but nothing substantial changes in the time evolution for anti-symmetric boundary conditions. The boundary at the large $m$ limit is trickier. At every time step, we must extrapolate extra lattice points outside the original lattice in order to evolve $f_m$ and $g_m$ at the boundary. Extrapolating 8 new lattice points at every time step makes the code unstable so a different approach is needed. An easy and working solution is to use linear approximation
\begin{align} f_m (t+\Delta t) &= f_m (t) + \Delta t\, F_m \big[\vec g\big], \\ g_m (t+\Delta t) &= g_m (t) - \Delta t\, F_m \big[\vec f\big] ,
\end{align}
for the lattice points $m_\text{max}-7,\ldots ,m_\text{max}$.
In doing so, one only needs to extrapolate two extra lattice points at the boundary which significantly reduces the numerical error. A simple linear extrapolation seems to be sufficient to obtain values for $f_m$ and $g_m$ on the two extra lattice points.

The reliability of the code was confirmed by checking energy conservation and unitarity of the time evolution, as well as by the apparent agreement with the semiclassical theory. To obtain a notion of energy describing the black hole mass, we calculate the expectation value of the Hamiltonian on the positive half line,
\begin{align} \label{eq:quantum_M}\langle M \rangle = \frac{1}{C^2 2\mu^{3/2}}\bigg[&\big(f_0^2+g_0^2\big) -(1+s) \big( f_0 f_2+ g_0 g_2\big) \nonumber \\ &+\sum_{m=1}^\infty \big(f_m^2+g_m^2\big) \Big( \sqrt{m-1}+\sqrt{m+1}+\mu^2 \sqrt{m}\Big) \nonumber \\ &-2 \sqrt{m+1}\Big( f_m f_{m+2}+ g_m g_{m+2}\Big) \bigg],
\end{align}
where the symmetry factor $s=1$ for symmetric boundary conditions and $s=-1$ for antisymmetric boundary conditions. The deviation of $\langle M \rangle$ from the classical value, $M=\sqrt{\phi_0}/2$, gives a quantitative measure of the extent to which the initial state is semiclassical. 
As one might expect, the initial Gaussian wave packet (\ref{eq:initial}) is semiclassical as long as the width of the state is large compared to the polymer scale and the wave amplitude is close to zero at the origin. In our investigations, the deviation from the classical mass was typically of the order of 0.5 \%, although deviations up to 1.5 \% were considered. In all these cases we found strong qualitative agreement with the effective polymer dynamics.

Numerical accuracy was monitored during each run in three ways. First the expectation value of the Hamiltonian was calculated at each time step. For all runs the relative change in $\langle M \rangle$ was kept between 0.1~\% and 0.15~\%. Typically this required time steps of the order of $1\times 10^{-7}$ units of Planck time. Secondly, the norm of the state was calculated at each time step. Using the time steps sizes as mentioned above, the relative error was also within 0.1~\% -- 0.15~\%, indicating the unitarity of the time evolution to the required order. Finally we checked on the convergence of the bounce value of $\langle \phi \rangle$ (denoted henceforth by $\phi_\text{min}$) with decreased time step. The convergence of $\phi_\text{min}$ in the limit $t\to 0$ was found to be almost linear, which made the error estimate rather easy. For the given sizes of the time steps, the estimate for relative error in $\phi_\text{min}$ was consistently below 0.15~\%, which is more than adequate for the purposes of our study. Smaller time steps would require significantly longer computation times and offer essentially no new information about the dynamics. 

\subsection{Results}
\begin{figure}[htbp!]
\begin{center}
\includegraphics[scale=0.6]{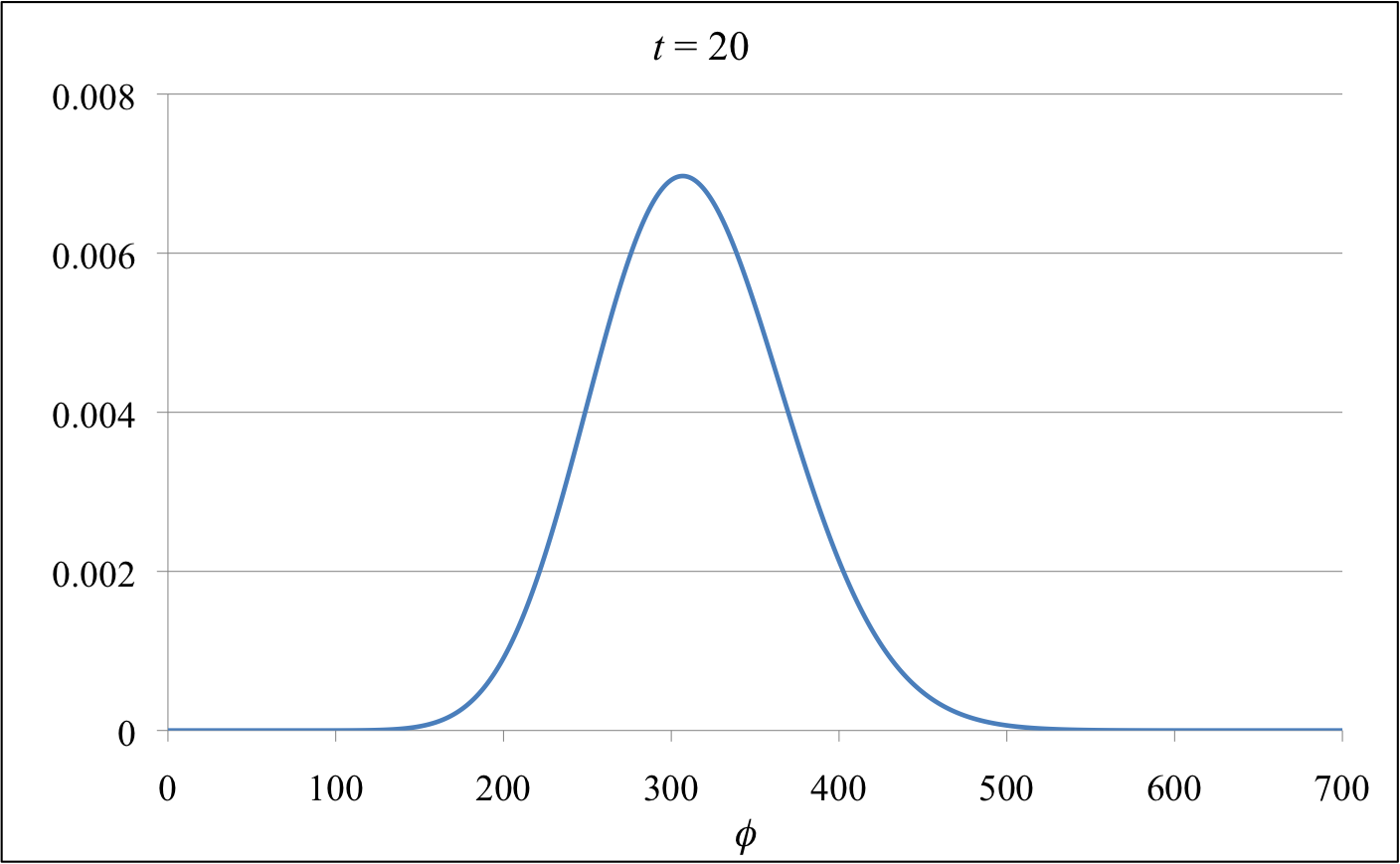}
\caption{Time evolution of wave packet, $t=20$. In this figure $\mu=1$, $\delta = 0.05$, $\phi_0=500$ and $\pi_0=-1\times 10^{-7}\approx 0$.}
\label{fig:phi500_t20}
\hfill

\includegraphics[scale=0.6]{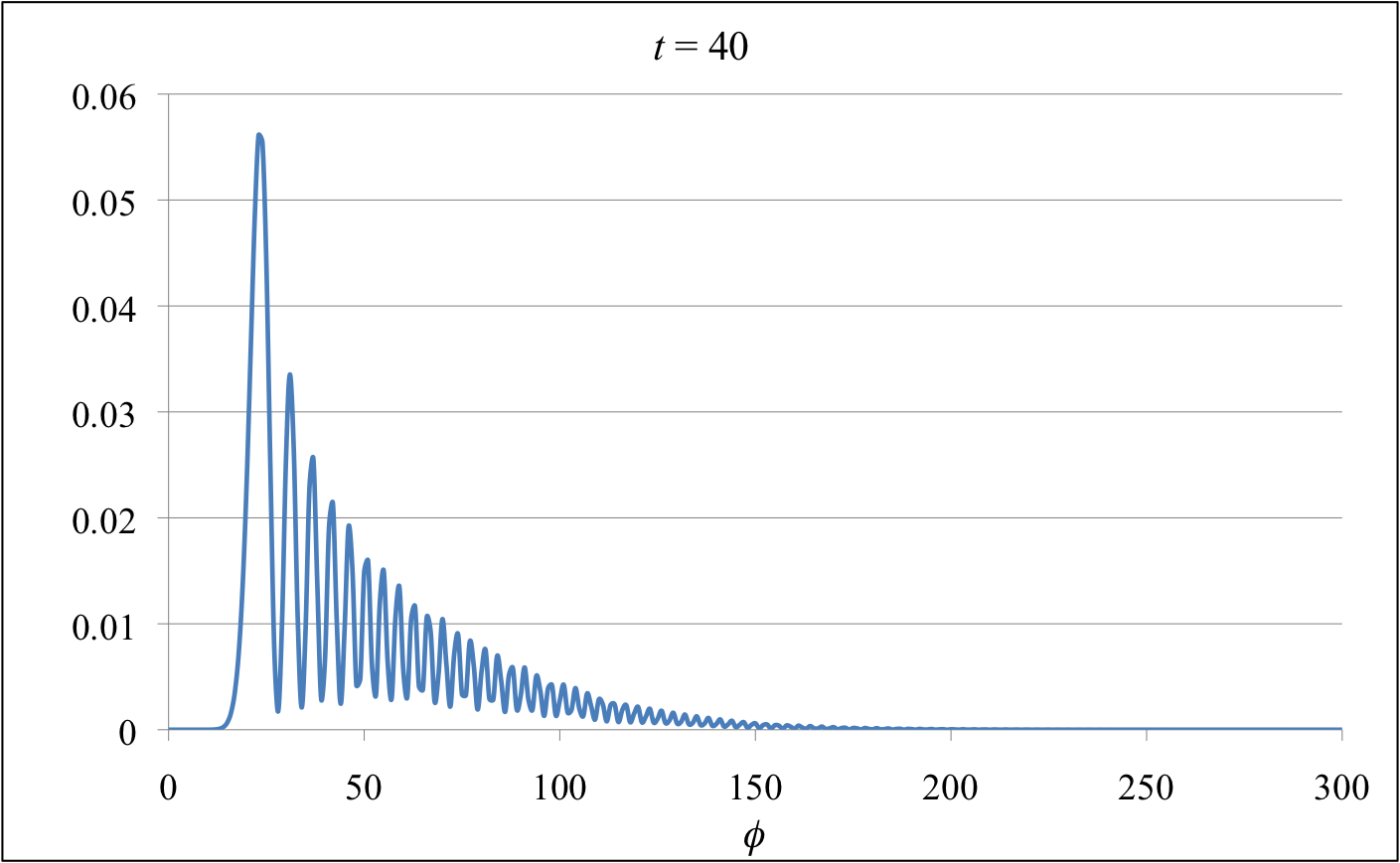}
\caption{Time evolution of wave packet, $t=35$. In this figure $\mu=1$, $\delta = 0.05$, $\phi_0=500$ and $\pi_0=-1\times 10^{-7}\approx 0$.}
\label{fig:phi500_t35}
\end{center}
\end{figure}

\begin{figure}[htbp!]
\begin{center}
\includegraphics[scale=0.6]{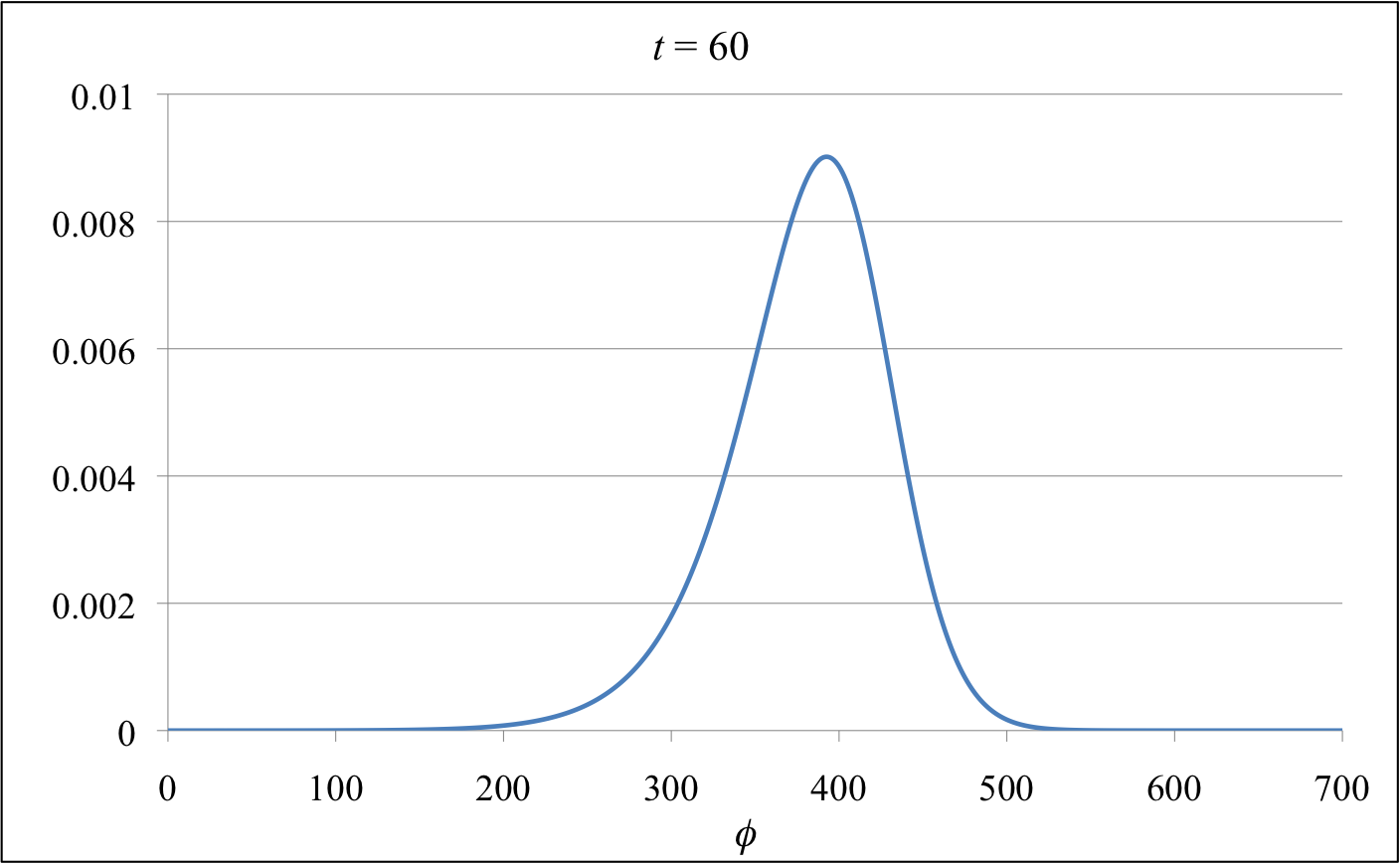}
\caption{Time evolution of wave packet, $t=60$. In this figure $\mu=1$, $\delta = 0.05$, $\phi_0=500$ and $\pi_0=-1\times 10^{-7}\approx 0$.}
\label{fig:phi500_t60}
\hfill

\includegraphics[scale=0.6]{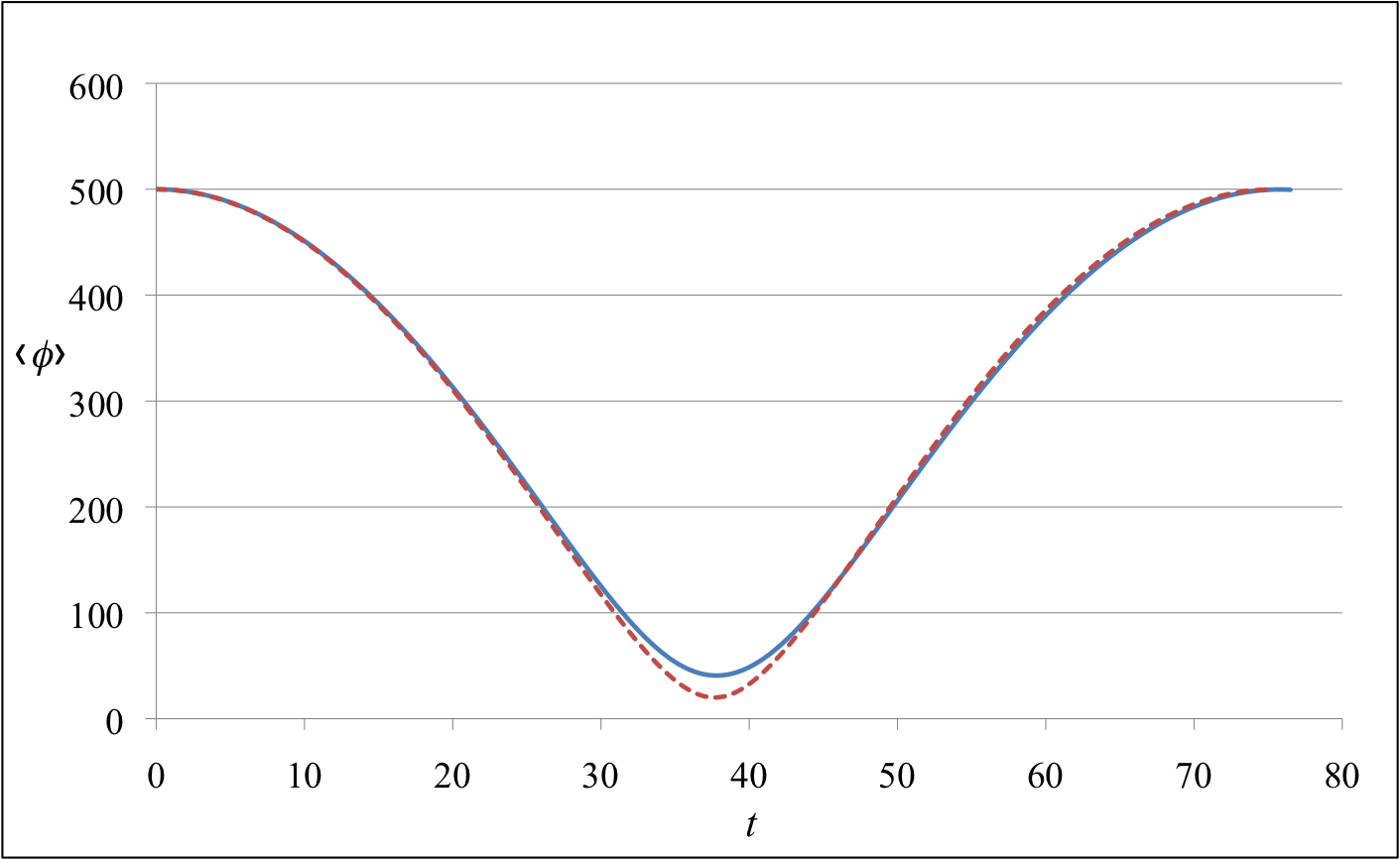}
\caption{Evolution of $\langle\phi\rangle$ as a function of proper time $t$. The top curve (blue, solid) shows is for the full polymer theory whereas the bottom curve (red, dashed) is for the effective theory. The former generally produces a bounce radius that is slightly larger than the latter.  In this figure $\mu=1$, $\delta = 0.05$, $\phi_0=500$ and $\pi_0=-1\times 10^{-7}\approx 0$.}
\label{fig:phi500}
\end{center}
\end{figure}

Figs.~\ref{fig:phi500_t0}-\ref{fig:phi500_t60} show the time evolution of a Gaussian pulse with symmetric boundary conditions. The initial expectation value of the throat area for this run is 500 $l_\text{Pl}^2$ with an initial pulse width of 14 $l_\text{Pl}^2$. The pulse initially spreads and then as it approaches the origin, it enters a quantum regime. As seen in Fig.~\ref{fig:phi500} the expectation value reaches its minimum value of about 41 Planck units at time $t\approx 38$. The time evolution of $\langle\phi\rangle$ is qualitatively in  agreement with the semiclassical estimate obtained from Eq.~(\ref{eq:t_phi}). A notable detail of the quantum evolution is that the maximum at $t\approx 76$ is slightly below $\phi_0 =500$.

The behavior of the width of the wave packet during the time evolution is somewhat unexpected. As seen from Fig.~\ref{fig:Delta_phi_b}, narrow initial width of the pulse does not necessarily imply narrow width at the bounce. In particular there exists a minimum value for $\Delta\phi$ at the bounce so that it cannot be made arbitrarily narrow by choosing the initial conditions. This appears to be a generic property for all initially Gaussian, semiclassical states. The width of the wave packet at the bounce (denoted henceforth by $\Delta \phi_b$) has significant effects on the nature of the bounce as described below. 

An interesting feature of the bounce is shown in Fig.~\ref{fig:phi500_ratio}. As the system approaches quantum regime, the {\it relative} spread of the wave packet $\Delta\phi/\langle\phi\rangle$ increases strongly, and at the bounce region it becomes of the order of unity.  For all the cases we were able to test, the maximum of $\Delta\phi/\langle\phi\rangle$ was roughly between 0.4 and 0.85, with the lowest values corresponding to the minimum of $\Delta\phi_b$ (cf. Fig.~\ref{fig:Delta_phi_b}). This means in particular that the uncertainty $\Delta \phi$ becomes comparable to the expectation value of $\phi$ at the bounce. Such strong fluctuations near the bounce naturally explain the relatively large values of the bounce radius that arise from the polymerized throat theory. It is also worth noticing that the relative spread of $\phi$ makes a tiny dip shortly after the bounce. This feature seems to persist regardless of the initial values of the parameters $\delta$, $\phi_0$ and~$\mu$. Ultimately the reason for this asymmetric form of the curve stems from the finite width of the wave packet: 
the natural tendency  to spread is balanced by the squeezing that occurs as the pulse moves towards the origin and occupies a smaller volume of configuration space.
This feature is not specific to the polymer theory but also arises in the Schr\"odinger quantization on the half-line. We have verified analytically that both free particle Gaussian wave packets and harmonic oscillator coherent states exhibit similar behavior for $\Delta\phi/\phi$ as the packet scatters off the origin. 

\begin{figure}[htbp!]
\begin{center}
\includegraphics[scale=0.6]{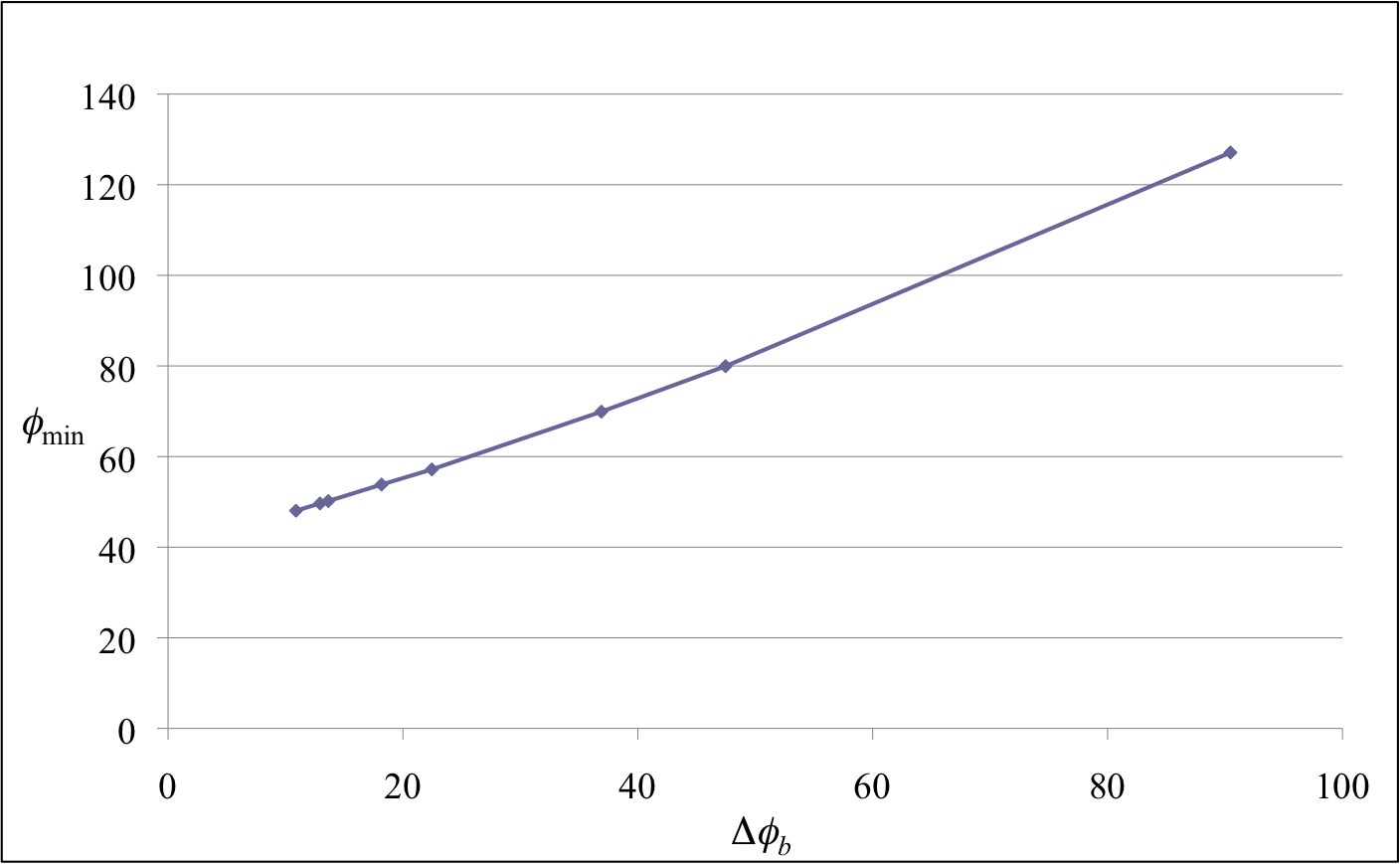}
\caption{The value of the minimum of $\langle\phi\rangle$ as a function of $\Delta\phi_b$. For all the data points $\mu = 1$, $\phi_0=1000$, and $\pi_0 = -1\times 10^{-7}\approx 0$. Although $\Delta \phi_b$ cannot be made arbitrary small in the full quantum theory, the figure strongly suggests that formally in the limit $\Delta \phi_b \rightarrow 0$, $\phi_\text{min}$ approaches the value predicted by the effective theory, which for given values of parameters is 40 $l^2_\text{Pl}$ (cf.\ Eq.~(\ref{eq:phi_-})).}
\label{fig:Delta2}
\hfill

\includegraphics[scale=0.6]{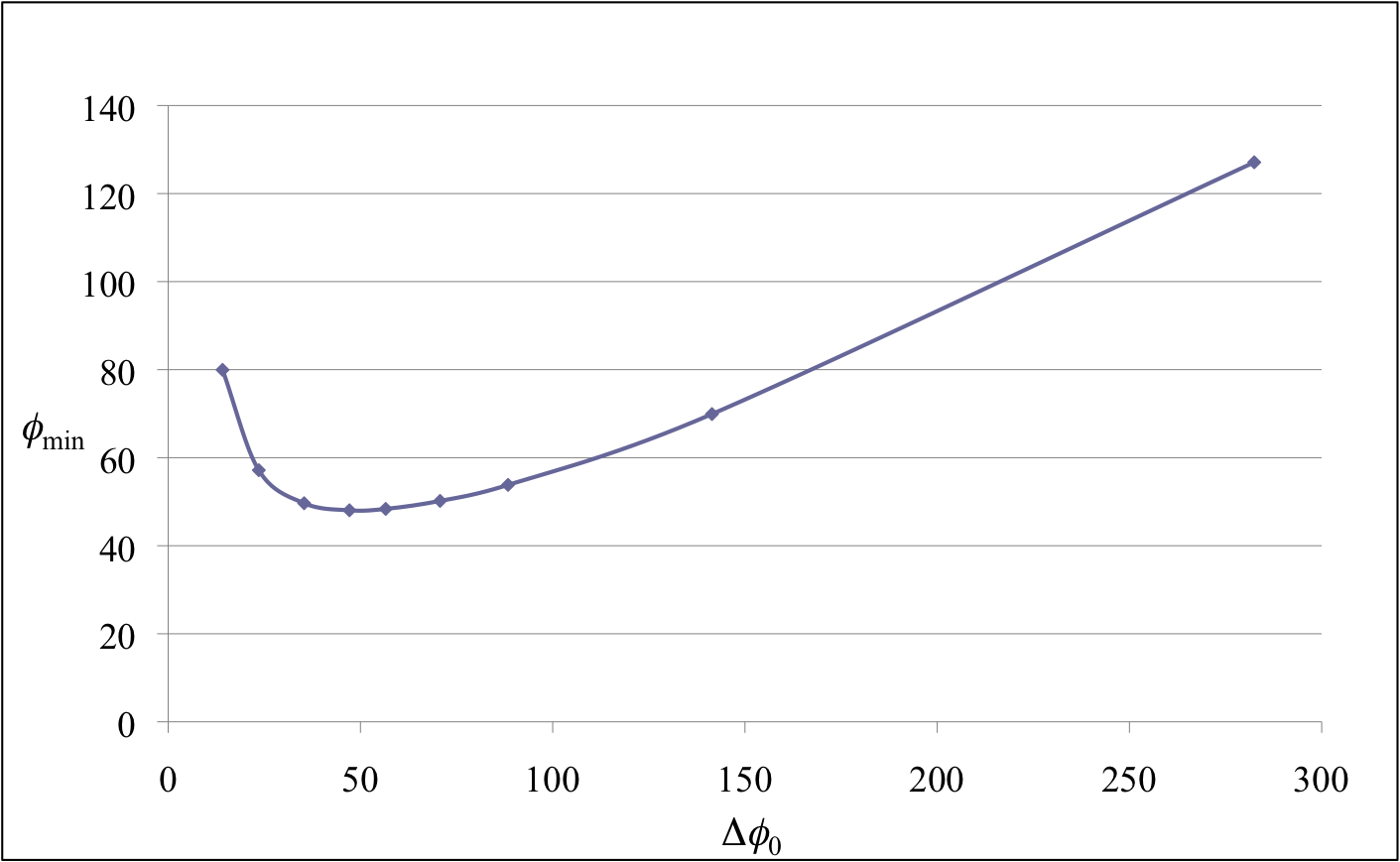}
\caption{The value of the minimum of $\phi_\text{min}$ as a function of~$\Delta\phi_\text{0}$. For all the data points $\mu = 1$, $\phi_0=1000$, and $\pi_0 = -1\times 10^{-7}\approx 0$.
\label{fig:phi_min_curve}}
\end{center}
\end{figure}

The quantum bounce area $\phi_\text{min}$ is generically somewhat higher than the semi-classical prediction. This also appears to be a consequence of the finite width of the wave packet at the  bounce. Fig.~\ref{fig:Delta2} shows how $\phi_\text{min}$ depends on the width $\Delta\phi_b$ of the pulse. As explained earlier, there is a limit of how narrow the pulse can be made at the bounce, and this, in turn, narrows the range of available data points. Nevertheless, it is quite clearly seen that in the limit where the width $\Delta\phi_b$ gets small, $\phi_\text{min}$ tends towards the value predicted by the effective theory, which for the initial parameters used in the figure equals 40 $l_\text{Pl}^2$ (cf.\ Eq.~(\ref{eq:phi_-})). Since the relationship between $\phi_\text{min}$ and $\Delta \phi_b$ is almost linear, it is no surprise that Fig.\ \ref{fig:phi_min_curve} of $\phi_\text{min}$ vs.\ $\Delta\phi_0$ is qualitatively similar to that of Fig.~\ref{fig:Delta_phi_b}. Our results provide numerical confirmation of the range of validity of the effective theory in the semi-classical limit. For finite width packets, one obtains a minimum area that is determined by the width and is more or less independent of the the polymerization scale. This confirms the expectation, based on the study of the Schr\"odinger quantization of the throat theory \cite{Louko:1996md}, that the central singularity in this model is also resolved in the continuum theory. 

\begin{figure}[htbp!]
\begin{center}
\includegraphics[scale=0.6]{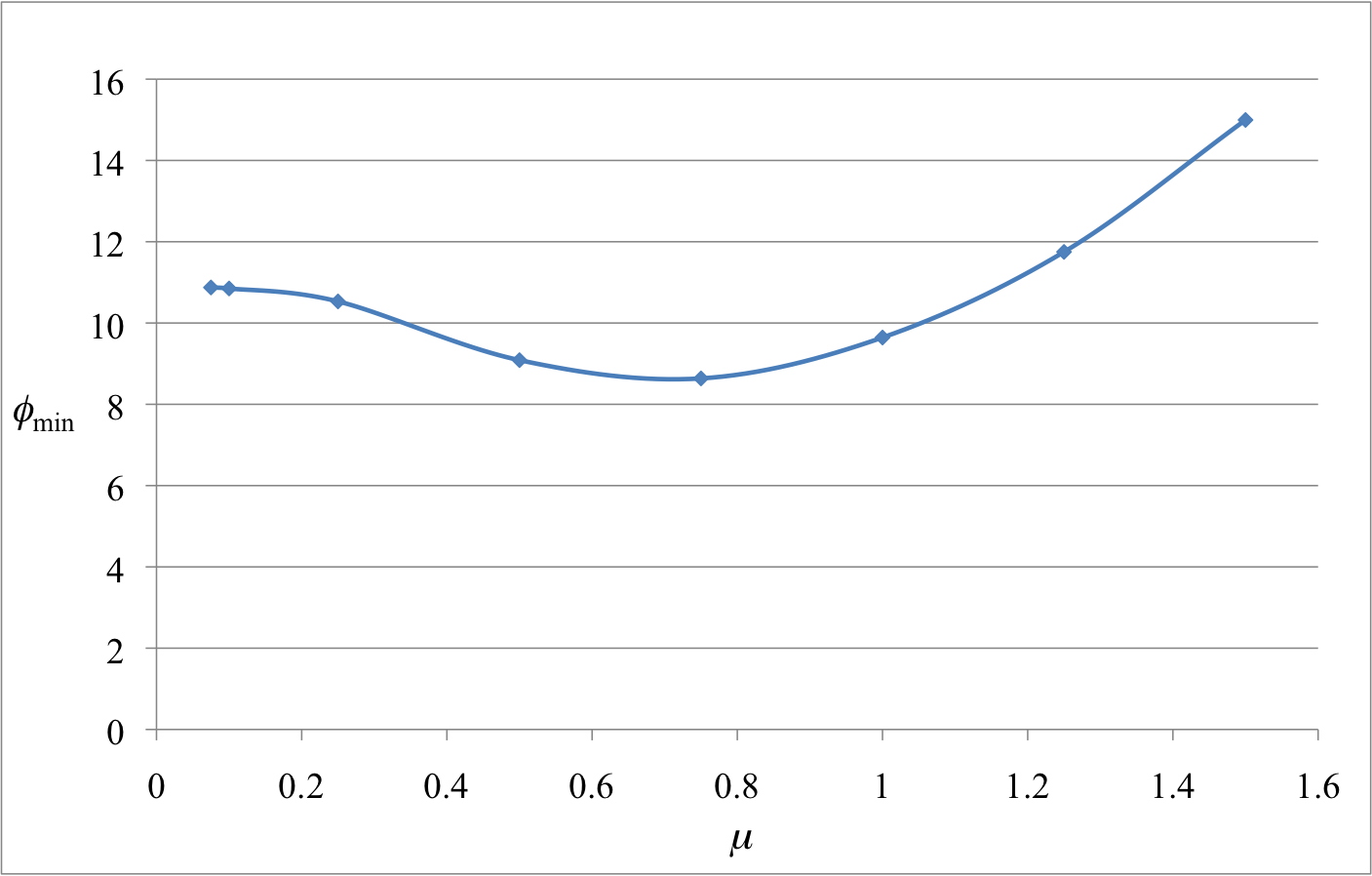}
\caption{The value of $\phi_\text{min}$ as a function of~$\mu$. All the data points correspond to the same the initial wave 
packet
with a physical width $\Delta \phi_0 \approx 17.7$, $\phi_0=80$, and $\pi_0 = -1\times 10^{-7}\approx 0$.
\label{fig:phi_vs_mu}}
\end{center}
\end{figure}

The singularity avoidance in the continuum limit may be seen in Fig.\ \ref{fig:phi_vs_mu} which shows the minimum area $\phi_\text{min}$ as a function of~$\mu$. The graph is obtained by keeping the physical width of the initial wave packet unchanged while decreasing the size of the lattice spacing $\mu$. As $\mu$ becomes smaller, the number of lattice points increases, making the numerics increasingly more difficult to handle. Using $\phi_0=80$ and $\Delta \phi_0 \approx 17.7$ we were able to get down to $\mu=0.075$ with reasonable computation times. With the given initial conditions, we see that the magnitude of $\phi_\text{min}$ does not change significantly when $\mu\lesssim 1$. We were also able to confirm that the time evolution of $\langle\phi\rangle$ is qualitatively similar to that of Fig.~\ref{fig:phi500} regardless of the size of~$\mu$. 
All this suggests that the existence of a finite polymer scale does  not significantly alter the qualitative aspects of the singularity resolution in this case.

\section{Conclusions}

We have considered the quantum dynamics of the Einstein-Rosen wormhole throat of an eternal Schwarzschild black hole using polymer quantization. An initially Gaussian, semi-classical wave packet remains semi-classical until it comes close to the singularity. The area expectation value reaches a positive minimum value as the packet is reflected from the singularity, and the packet emerges from the bounce to form a new, near-Gaussian semi-classical state. The bounce area depends in a non-trivial way on the initial energy and width of the pulse, and also on the polymerization scale: however, for wave packets that remain sufficiently narrow near the bounce, we find that the bounce area agrees with that of the semiclassical polymer theory~\cite{kpl}. 

Although the numerics become difficult to control as the polymerization scale decreases, our results provide evidence that for pulses of finite width the bounce persists as the polymerization scale keeps decreasing. 
We have not investigated numerically the behaviour of wave packets in the corresponding Schr\"odinger theory~\cite{Louko:1996md,kpl}, but the unitarity of the Schr\"odinger theory implies that a bounce of some sort must occur, and we are not aware of reasons to expect that the small polymerization scale limit of the polymer bounce would differ from the Schr\"odinger bounce. 

\begin{figure}[htbp!]
\begin{center}
\includegraphics[scale=0.6]{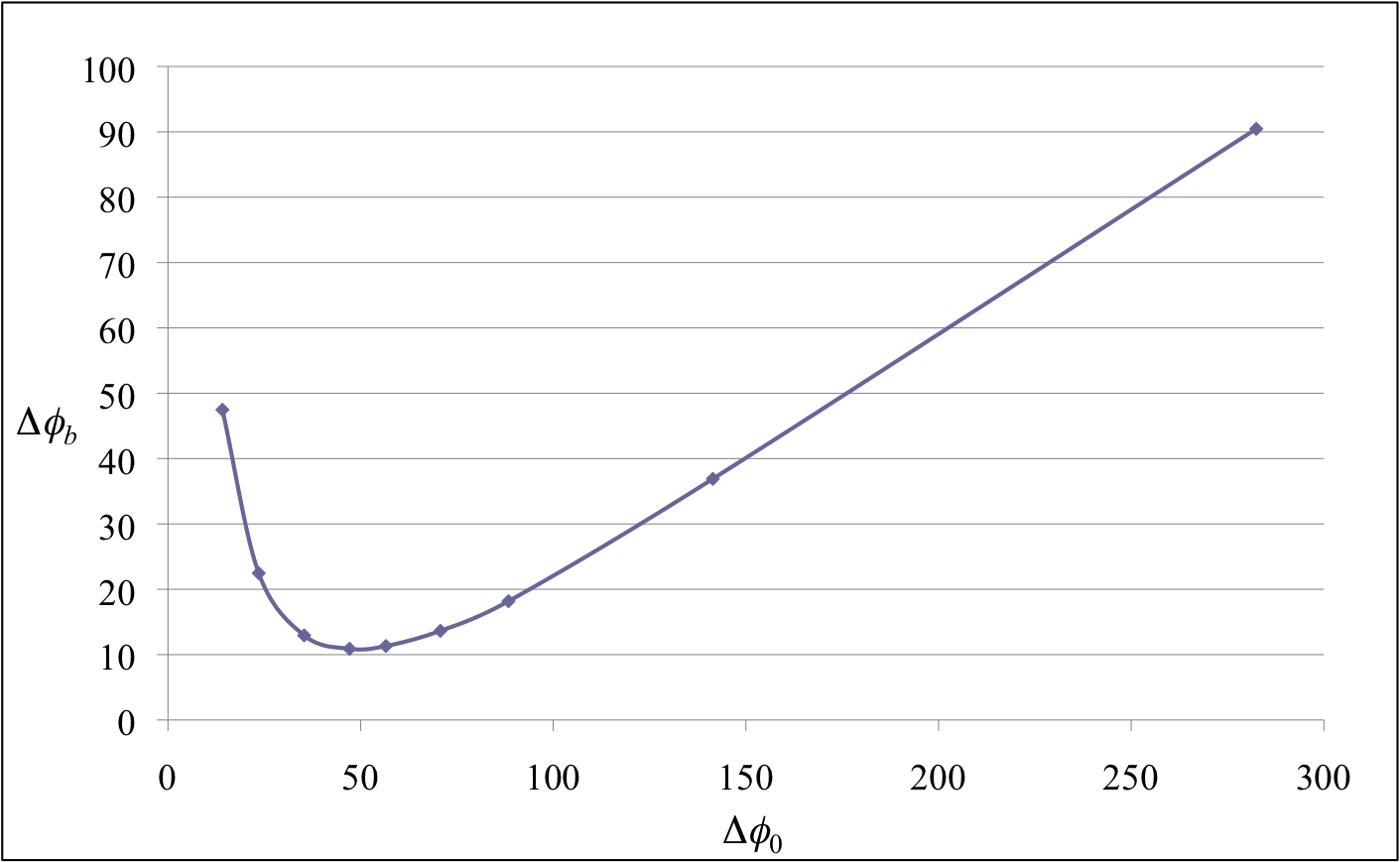}
\caption{The value of the standard deviation of $\phi$ at the bounce, $\Delta \phi_b$, as a function of the initial value of the standard deviation $\Delta \phi_0$. In this figure $\mu=1$, $\phi_0=1000$ and $\pi_0=-1\times 10^{-7}\approx 0$.}
\label{fig:Delta_phi_b}
\hfill

\includegraphics[scale=0.6]{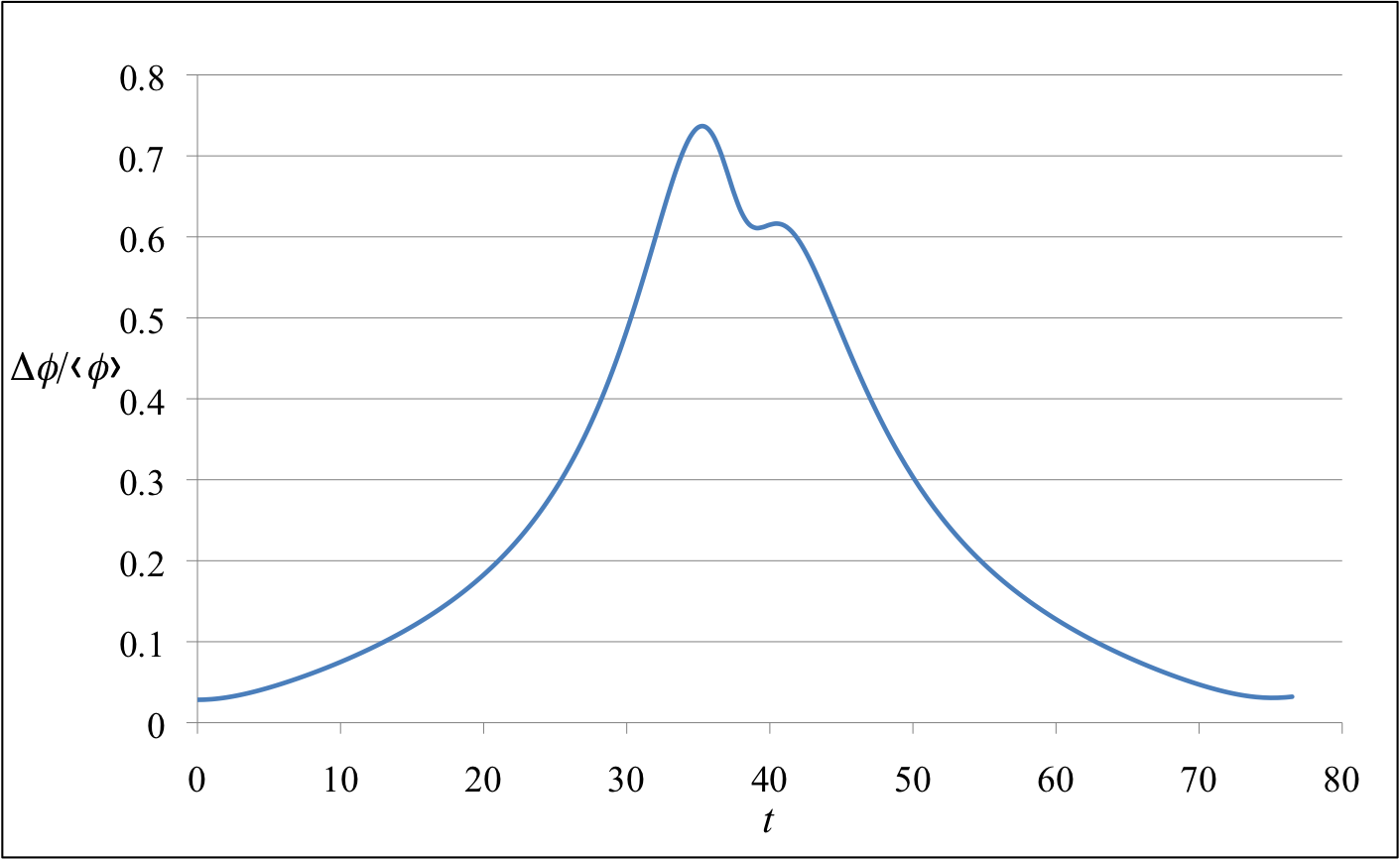}
\caption{The value of $\Delta\phi/\langle\phi\rangle$ as a function of the time.  In this figure $\mu=1$, $\delta = 0.05$, $\phi_0=500$ and $\pi_0=-1\times 10^{-7}\approx 0$.}
\label{fig:phi500_ratio}
\end{center}
\end{figure}

For our wave packets, the wormhole area expectation value $\langle \phi \rangle$ is almost time-symmetric about the bounce, but the relative uncertainty $\Delta \phi / \langle \phi \rangle$ is not, as seen in Figure~\ref{fig:phi500_ratio}. This asymmetry is a consequence of the spreading of the wave packet as it evolves. We have verified analytically that a similar asymmetry occurs also in Schr\"odinger quantization of a free point particle on the positive half-line with the corresponding boundary condition. In fact, for the free point particle the spreading is so strong that the packet must be carefully tuned to be narrow near the bounce in order to see a significant peaking in $\Delta \phi / \langle \phi \rangle$ about the bounce at all. 

Overall, our results do not suggest significant qualitative differences between the polymer and Schr\"odinger singularity avoidance mechanisms for the Einstein-Rosen wormhole throat. In this respect the singularity of the throat theory appears to be no more insidious than that of the Coulomb potential~\cite{hlw}. We note that while the classical throat theory is completely deparametrized and involves no constraints, differences in polymer and continuum singularity resolution do arise in symmetry-reduced systems where constraints are present 
\cite{lqc1,lqc2,lqc3,Ashtekar:2007em,%
corichi08,ashtekar05,modesto06,boehmer07,pullin08,nelson08,pk09}; 
see in particular \cite{Ashtekar:2007em} for a situation where continuum quantization is argued to lead to physically unsatisfactory results. 
This suggests that the differences between polymer and Schr\"odinger quantizations may relate less to the singularities of the underlying classical theory than to whether the constraints are addressed prior to or after the polymer structure is introduced. 

\vspace{3ex}

\par\noindent
{\bf Acknowledgements:} 
GK and AP were supported in part by the Natural Sciences and 
Engineering Research Council of Canada. 
JL~was supported in part by STFC (UK) grant PP/D507358/1.

\newpage


\end{document}